\documentclass[a4paper,11pt]{article}
\usepackage{jheppub}

\title{\boldmath Configurational Temperature in Matrix Models and Random Matrix Ensembles}

\author[a,b]{Anosh Joseph}
\author[a]{and Vinod Mamale}

\affiliation[a]{National Institute for Theoretical and Computational Sciences, \\ School of Physics, and Mandelstam Institute for Theoretical Physics,\\ University of the Witwatersrand, Johannesburg, Wits 2050, South Africa}
\affiliation[b]{Brown Center for Theoretical Physics and Innovation, \\Department of Physics, Brown University, \\ Providence, RI 02912, United States}

\emailAdd{anosh.joseph@wits.ac.za}
\emailAdd{vinod.mamale@wits.ac.za}

\abstract{
We investigate the configurational temperature estimator in interacting matrix models and Gaussian random-matrix ensembles. The estimator follows from an exact Schwinger--Dyson identity and may be expressed in terms of the gradient and Hessian of the action. We study the Gross--Witten--Wadia model, a quartic double-well matrix model, and the Gaussian Orthogonal, Unitary, and Symplectic Ensembles. In all cases, the estimator satisfies the exact Schwinger--Dyson identity, $\beta_{\rm config} = 1$, within statistical uncertainties. Separating the estimator into isotropic and anisotropic parts, we find that the leading finite-$N$ corrections satisfy the approximate relation $\beta_{\rm iso} - 1 \simeq - \beta_{\rm aniso}$. We also show that the configurational temperature estimator provides a sensitive diagnostic of Monte Carlo simulations.
}

\begin{document}
\maketitle
\flushbottom

\section{Introduction}
\label{sec:Introduction}

Monte Carlo methods play a central role in the nonperturbative study of statistical systems, quantum field theories, and matrix models. 
Their reliability ultimately depends on the ability of the simulation algorithm to sample the desired probability measure accurately. 
In practice, this is commonly assessed by measuring selected observables and monitoring equilibration, autocorrelations, and acceptance rates. 
While these diagnostics are often effective, they typically probe only specific aspects of the sampled distribution and may not directly test whether the underlying probability measure has been reproduced correctly.

Configurational temperature provides an alternative perspective. 
Originally introduced in the statistical-mechanics literature as a temperature estimator constructed from derivatives of the potential energy \cite{Rugh:1997, Rugh:1998, Butler:1998, Jepps:2000}, it can be understood more generally as a consequence of an exact integration-by-parts identity of the probability measure \cite{Joseph:2026hom}. 
In Euclidean path integrals and matrix models, this identity takes the form of a Schwinger--Dyson relation involving the gradient and Hessian of the action. 
As a result, configurational temperature defines an exact observable whose expectation value is fixed by the probability measure itself. 

The configurational-temperature construction is appealing for several reasons. 
First, it depends only on derivatives of the action and therefore can be evaluated directly from the sampled configurations. 
Second, because it follows from an exact Schwinger--Dyson identity, it provides a stringent consistency condition for numerical simulations. 
Finally, its representation in terms of the gradient and Hessian naturally probes local properties of the probability measure and therefore contains information that is complementary to conventional observables. 

In this work, we investigate configurational temperature in a variety of matrix models and random-matrix ensembles. 
Our primary focus is on the Gross--Witten--Wadia (GWW) matrix model \cite{Gross:1980he, Wadia:1980cp, Wadia:2012fr} and a quartic double-well matrix model, which provide examples of interacting systems with nontrivial eigenvalue dynamics. 
To place these results in a broader context, we also study the Gaussian Orthogonal, Unitary, and Symplectic Ensembles (GOE, GUE, and GSE). 
These ensembles possess comparatively simple equilibrium measures while retaining the nontrivial eigenvalue correlations induced by the Vandermonde determinant.

A central goal of this work is to explore how configurational temperature approaches its exact value in finite-$N$ systems. (See Refs. \cite{Dhindsa:2025xfv, Longia:2026doi} for recent results from U(1) lattice gauge theories.)
We show that the estimator admits a natural decomposition into isotropic and anisotropic sectors. 
Although the sum of these contributions satisfies the exact identity
\begin{equation}
\beta_{\rm config} = 1,
\end{equation}
the individual sectors exhibit nontrivial finite-$N$ corrections. 
By studying these corrections across several ensembles, we obtain a quantitative characterization of the large-$N$ behavior of the configurational-temperature estimator.

Our numerical results reveal a common pattern. 
In all models considered, the isotropic and anisotropic sectors exhibit finite-$N$ corrections of comparable magnitude and satisfy the approximate relation
\begin{equation}
\beta_{\rm iso}-1 \simeq -\beta_{\rm aniso}.
\end{equation}
The approximate cancellation between the leading finite-$N$ corrections is consistent with the exact configurational-temperature identity while generating a nontrivial finite-$N$ structure. 
The associated scaling exponents depend on the ensemble under consideration, ranging from approximately $-0.35$ in the GOE to approximately $-0.87$ in the GSE, with the interacting matrix models yielding intermediate values near $-0.6$ and $-0.65$.

A second objective of this work is to explore configurational temperature as a practical diagnostic of Monte Carlo simulations. 
Because its expectation value is fixed by an exact identity of the target probability measure, deviations from the exact identity provide a direct signal of sampling errors. 
We demonstrate that the estimator can detect incomplete thermalization and systematic distortions of the probability measure even in situations where conventional observables appear comparatively insensitive. 
Configurational temperature has recently been proposed as a diagnostic of sampling correctness in Complex Langevin simulations and related stochastic sampling methods
\cite{Joseph:2025fcd, Joseph:2025xbn, Joseph:2026xti}.

The remainder of this paper is organized as follows. 
In Sec.~\ref{sec:ConfigTemp} we derive the configurational temperature estimator and introduce its decomposition into isotropic and anisotropic sectors. 
Section~\ref{sec:Models} presents the matrix models and numerical methods used in this work. 
The behavior of configurational temperature in the interacting matrix models is examined in Sec.~\ref{sec:MatrixResults}, while Sec.~\ref{sec:GaussianResults} contains the corresponding analysis for the Gaussian random-matrix ensembles. 
In Sec.~\ref{sec:Diagnostics}, we investigate configurational temperature as a diagnostic of Monte Carlo simulations. Section~\ref{sec:Discussion} contains a discussion of the results and possible future directions. 

\section{Configurational Temperature} 
\label{sec:ConfigTemp} 

Configurational temperature is an observable constructed directly from the probability measure. 
Unlike conventional observables, which probe specific moments or correlation functions, it depends on derivatives of the action and therefore provides information about the local structure of the sampled distribution. 
Because it follows from an exact Schwinger--Dyson identity, configurational temperature provides a useful consistency condition for Monte Carlo simulations. 
Its representation in terms of the gradient and Hessian of the action also permits a geometric interpretation in configuration space.

In this section, we derive the configurational-temperature estimator and introduce a decomposition into isotropic and anisotropic contributions. 
This decomposition will serve as the basis for the numerical investigations presented later in the paper. 

\subsection{Schwinger--Dyson Identity} 

Consider a system with configuration-space coordinates
\begin{equation}
\phi = (\phi_1, \phi_2, \ldots, \phi_D) 
\end{equation} 
distributed according to the probability measure 
\begin{equation} 
P[\phi] = \frac{1}{Z} e^{-S[\phi]}, \qquad Z = \int d^D\phi \, e^{-S[\phi]}. 
\end{equation} 

For simplicity we present the derivation for a finite-dimensional configuration space. 
The corresponding functional Schwinger--Dyson identity for Euclidean field theories follows formally by replacing ordinary derivatives with functional derivatives and the measure $d^D \phi$ with the functional measure ${\cal D} \phi$.

Expectation values are defined by 
\begin{equation} 
\langle O \rangle = \frac{ \int d^D\phi \, O[\phi] \, e^{-S[\phi]} }{ \int d^D\phi \, e^{-S[\phi]} }. 
\end{equation} 

Let 
\begin{equation}
F(\phi) = (F_1, \ldots, F_D) 
\end{equation} 
be a sufficiently smooth vector field on configuration space. 
Assuming the probability measure and vector field are such that boundary terms vanish, integration by parts gives 
\begin{equation} 
\int d^D\phi \, \frac{\partial}{\partial\phi_i} \Big( F_i e^{-S} \Big) = 0. 
\end{equation} 

Expanding the derivative yields 
\begin{equation} 
\int d^D\phi 
\left( \frac{\partial F_i}{\partial\phi_i} - F_i \frac{\partial S}{\partial\phi_i} \right) e^{-S} = 0. 
\end{equation} 

Expressed in vector notation, this relation becomes 
\begin{equation} 
\left\langle \nabla \cdot F \right\rangle = \left\langle F \cdot \nabla S \right\rangle. 
\label{eq:generalSD} 
\end{equation} 

This expression is valid for any sufficiently smooth vector field $F$ for which boundary contributions vanish. 

Equation~(\ref{eq:generalSD}) is a Schwinger--Dyson identity associated with the infinitesimal transformation 
\begin{equation} 
\phi_i \rightarrow \phi_i + \epsilon F_i(\phi). 
\end{equation} 
Different choices of the vector field $F$ generate different exact identities of the probability measure. 

\subsection{Configurational Temperature} 

A particularly useful choice is a gradient field of the form
\begin{equation} 
F = \frac{\nabla S} {\nabla S\cdot\nabla S}. 
\end{equation} 
For this vector field, one has 
\begin{equation} 
F \cdot \nabla S = 1, 
\end{equation} 
and Eq.~(\ref{eq:generalSD}) reduces to 
\begin{equation} 
\left\langle \nabla \cdot \left( \frac{\nabla S} {\nabla S \cdot \nabla S} \right) \right\rangle = 1. 
\end{equation} 

This motivates the definition of the configurational temperature, 
\begin{equation} 
\beta_{\rm config} \equiv \left\langle \nabla \cdot \left( \frac{\nabla S} {\nabla S \cdot \nabla S} \right) \right\rangle. 
\label{eq:betaConfigDef} 
\end{equation} 

For any ensemble distributed according to the target probability measure,
\begin{equation} 
\beta_{\rm config} = 1. 
\label{eq:betaConfigIdentity} 
\end{equation} 

Equation~(\ref{eq:betaConfigIdentity}) is an exact consequence of the probability measure and is independent of the details of the underlying model. 
In practical applications, deviations from unity signal that the sampled distribution differs from the target measure. 
Throughout this work, we use Eq.~(\ref{eq:betaConfigIdentity}) both as a consistency check and as a starting point for investigating the finite-$N$ behavior of the estimator. 

\subsection{Gradient--Hessian Representation} 

To obtain a more explicit form, we introduce the gradient 
\begin{equation} 
g_i = \frac{\partial S}{\partial\phi_i}, 
\end{equation} 
and the Hessian 
\begin{equation} 
H_{ij} = \frac{\partial^2 S}{\partial \phi_i \partial \phi_j}. 
\end{equation} 

Defining 
\begin{equation} 
|g|^2 = \sum_i g_i^2, 
\end{equation} 
a straightforward calculation gives 
\begin{equation} 
\nabla \cdot \left( \frac{g}{|g|^2} \right) = \frac{{\rm Tr} \, H}{|g|^2} - 2 \, \frac{g^T H g}{|g|^4}. 
\end{equation} 

The configurational temperature may therefore be written as 
\begin{equation} 
\beta_{\rm config}  = \left\langle \frac{{\rm Tr} \, H}{|g|^2} - 2 \, \frac{g^T H g}{|g|^4} \right\rangle. 
\label{eq:betaConfigGH} 
\end{equation} 

Equation~(\ref{eq:betaConfigGH}) expresses the configurational temperature entirely in terms of the first and second derivatives of the action. 
This representation is particularly convenient for matrix models, where both the gradient and Hessian can be computed explicitly in the eigenvalue representation. 

\subsection{Isotropic and Anisotropic Sectors} 

The two terms appearing in Eq.~(\ref{eq:betaConfigGH}) naturally define a decomposition 
\begin{equation} 
\beta_{\rm config} = \beta_{\rm iso} + \beta_{\rm aniso}, 
\label{eq:betaDecomposition} 
\end{equation} 
with 
\begin{equation} 
\beta_{\rm iso} = \left\langle \frac{{\rm Tr} \, H}{|g|^2} \right\rangle, 
\label{eq:betaIso} 
\end{equation} 
and 
\begin{equation} 
\beta_{\rm aniso} = - 2 \left\langle \frac{g^T H g}{|g|^4} \right\rangle. 
\label{eq:betaAniso} 
\end{equation} 

The quantity $\beta_{\rm iso}$ depends on the trace of the Hessian and therefore probes curvature averaged over all directions in configuration space. 
We refer to this contribution as the isotropic sector. 

The second contribution depends on the projection of the Hessian along the gradient direction. 
Introducing the normalized gradient 
\begin{equation} 
\hat g = \frac{g}{|g|}, 
\end{equation} 
one may write 
\begin{equation}
\frac{g^T H g}{|g|^2} = \hat g^T H \hat g. 
\end{equation} 
The anisotropic sector, therefore, probes the projection of the Hessian along the local gradient direction. 

The decomposition \eqref{eq:betaDecomposition} separates the configurational-temperature estimator into a contribution associated with the trace of the Hessian and a contribution associated with its projection along the gradient direction.

In the following sections, we examine the behavior of these quantities in a variety of matrix models and random-matrix ensembles, with particular emphasis on their finite-$N$ scaling properties.

\section{Matrix Models and Numerical Setup} 
\label{sec:Models} 

In this section, we introduce the matrix models studied in this work and summarize the numerical methods used in our simulations. 
Throughout, our analysis is performed in the eigenvalue representation, where the configurational-temperature estimator may be evaluated directly from the gradient and Hessian of the effective action. 

The models considered here were chosen to provide a range of probability measures with different interaction structures. 
The Gross--Witten--Wadia and double-well models represent interacting matrix theories with nontrivial large-$N$ dynamics, while the Gaussian ensembles provide analytically well-understood reference systems.

\subsection{Gross--Witten--Wadia Model} 
\label{sec:GWW} 

The Gross--Witten--Wadia (GWW) model is one of the simplest examples of a unitary matrix model with a large-$N$ phase transition. 
It is defined in terms of an $N \times N$ unitary matrix $U \in U(N)$ and a coupling constant $\lambda$. 

The partition function is
\begin{equation}
Z[\lambda] = \int_{U(N)} dU ~ e^{ - S[U] } = \int_{U(N)} dU ~ \exp \left[ - \frac{N}{2 \lambda} \left( {\rm Tr} U + {\rm Tr} U^\dagger \right) \right].
\end{equation}

Diagonalizing the unitary matrix according to 
\begin{equation} 
U = {\rm diag} \left( e^{i\theta_1}, e^{i\theta_2}, \ldots, e^{i\theta_N} \right), 
\end{equation} 
the partition function may be written as 
\begin{equation} 
Z = \int \prod_{i = 1}^{N} d\theta_i \, e^{-S(\theta)}, 
\end{equation} 
with effective action 
\begin{equation} 
S(\theta) = - \frac{N}{\lambda} \sum_{i = 1}^{N} \cos\theta_i - \sum_{i < j} \log \left[ \sin^2 \left( \frac{\theta_i - \theta_j}{2} \right) \right]. 
\label{eq:GWWaction}
\end{equation} 

The gradient entering the configurational-temperature estimator is 
\begin{equation} 
g_i = \frac{\partial S} {\partial\theta_i} = \frac{N}{\lambda} \sin\theta_i - \sum_{j \neq i} \cot \left( \frac{\theta_i - \theta_j}{2} \right), 
\end{equation} 
while the Hessian matrix is given by 
\begin{equation} 
H_{ii} = \frac{N}{\lambda} \cos\theta_i + \frac12 \sum_{j \neq i} \csc^2 \left( \frac{\theta_i - \theta_j}{2} \right), 
\end{equation} 
and 
\begin{equation} 
H_{ij} = - \frac12 \csc^2 \left( \frac{\theta_i - \theta_j}{2} \right), \qquad i \neq j. 
\end{equation} 

\subsection{Double-Well Matrix Model} 
\label{sec:DW} 

The second interacting system considered in this work is the quartic double-well matrix model, 
\begin{equation} 
Z = \int dM \, \exp \left[ - N \, {\rm Tr} \left( - \frac{\mu}{2} M^2 + \frac14 M^4 \right) \right], 
\label{eq:DWpartition} 
\end{equation} 
where $M$ is an $N \times N$ Hermitian matrix and $\mu$ controls the shape of the potential. 

After diagonalization, 
\begin{equation} 
M = U \Lambda U^\dagger, \qquad \Lambda = {\rm diag} (\lambda_1, \ldots, \lambda_N), 
\end{equation} 
the partition function becomes 
\begin{equation} 
Z = \int \prod_{i = 1}^{N} d\lambda_i \, e^{-S(\lambda)}, 
\end{equation} 
with effective action 
\begin{equation} 
S(\lambda) = N \sum_{i = 1}^{N} \left( - \frac{\mu}{2} \lambda_i^2 + \frac14 \lambda_i^4 \right) - 2 \sum_{i < j} \log |\lambda_i - \lambda_j|. 
\label{eq:DWaction} 
\end{equation} 

The gradient is 
\begin{equation} 
g_i = N \left( - \mu \lambda_i + \lambda_i^3 \right) - 2 \sum_{j \neq i} \frac{1}{\lambda_i - \lambda_j}, 
\end{equation} 
and the Hessian is 
\begin{equation} 
H_{ii} = N \left( - \mu + 3 \lambda_i^2 \right) + 2 \sum_{j \neq i} \frac{1}{(\lambda_i - \lambda_j)^2}, 
\end{equation} 
\begin{equation} 
H_{ij} = - \frac{2} {(\lambda_i - \lambda_j)^2}, \qquad i \neq j. 
\end{equation}  

\subsection{Gaussian Ensembles} 
\label{sec:GaussianModels} 

To provide a reference point for the interacting matrix models, we also consider the classical Gaussian random-matrix ensembles. 
These may be treated simultaneously by introducing the Dyson index 
\begin{equation} 
\gamma = \begin{cases} 
1, & {\rm GOE}, \\ 
2, & {\rm GUE}, \\ 
4, & {\rm GSE}. 
\end{cases} 
\end{equation} 

In the eigenvalue representation, the partition function takes the form 
\begin{equation} 
Z = \int \prod_{i = 1}^{N} d \lambda_i \, e^{-S(\lambda)}, 
\end{equation} 
with effective action 
\begin{equation} 
S(\lambda) = \frac{a}{2} \sum_{i = 1}^{N} \lambda_i^2 - \gamma \sum_{i < j} \log|\lambda_i - \lambda_j|. 
\label{eq:GaussianAction} 
\end{equation} 
The parameter $a$ fixes the overall scale of the Gaussian potential.
Throughout this work, we set $a = 1$, unless otherwise stated.

The corresponding gradient and Hessian are 
\begin{equation} 
g_i = a \lambda_i - \gamma \sum_{j \neq i} \frac{1}{\lambda_i - \lambda_j}, 
\end{equation} 
\begin{equation} 
H_{ii} = a + \gamma \sum_{j \neq i} \frac{1}{(\lambda_i - \lambda_j)^2}, 
\end{equation} 
and 
\begin{equation} 
H_{ij} = - \frac{\gamma} {(\lambda_i - \lambda_j)^2}, \qquad i \neq j. 
\end{equation} 

These expressions provide all ingredients required to evaluate the configurational-temperature estimator and its decomposition into isotropic and anisotropic sectors.

\subsection{Numerical Setup} 
\label{sec:Numerics} 

The numerical results presented in this work were obtained using standard Metropolis Monte Carlo simulations in the eigenvalue representation (see, e.g., Ref.~\cite{Joseph:2019zer}).
For each model, candidate configurations were generated via local updates to the eigenvalues, with acceptance determined by the resulting change in the effective action. 

For every configuration, we evaluated the configurational-temperature estimator 
\begin{equation} 
\beta_{\rm config} = \beta_{\rm iso} + \beta_{\rm aniso} 
\end{equation} 
using the gradient and Hessian expressions given above. 
Ensemble averages were computed after discarding an initial thermalization period.  

We characterize the observed finite-$N$ behavior by fitting the data to the empirical power-law form
\begin{equation}
A N^B,
\end{equation}
where both the amplitude $A$ and exponent $B$ were treated as free parameters.

In the following sections, we present the resulting measurements of the configurational-temperature estimator and its isotropic and anisotropic components in the interacting matrix models and Gaussian ensembles. 

\section{Configurational Temperature in Matrix Models} 
\label{sec:MatrixResults} 

In this section, we present numerical measurements of the configurational-temperature estimator and its decomposition into isotropic and anisotropic sectors in the Gross--Witten--Wadia and double-well matrix models. 
We first verify the exact identity $\beta_{\rm config} = 1$, and then investigate the finite-$N$ behavior of the individual contributions 
\begin{equation} 
\beta_{\rm config} = \beta_{\rm iso} + \beta_{\rm aniso}. \nonumber
\end{equation} 

Particular attention will be paid to the scaling of the isotropic and anisotropic sectors with matrix size and to the extent to which their finite-$N$ corrections exhibit similar behavior. 

\subsection{Verification of the Configurational-Temperature Identity} 

\begin{figure*}[htbp]
\centering
\includegraphics[width=14cm]{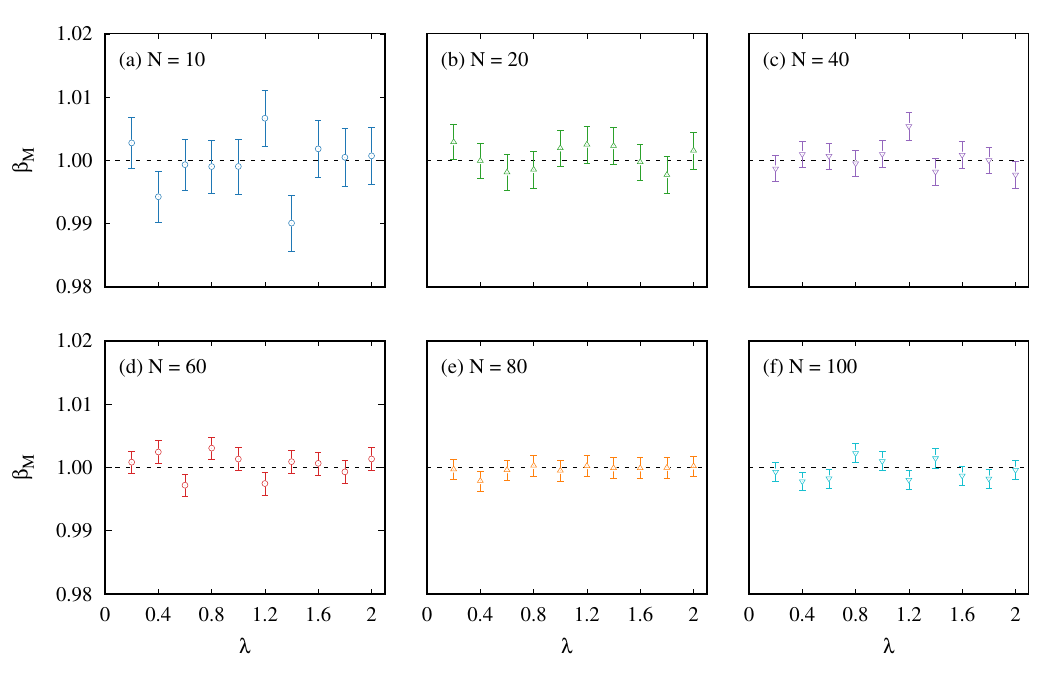}
\caption{The Gross--Witten--Wadia model. Measured configurational temperature $\beta_M$ as a function of the 't Hooft coupling $\lambda$ for $N = 10, 20, 40, 60, 80,$ and $100$. Across the full coupling range studied, the estimator remains consistent with the exact Schwinger--Dyson identity $\beta_{\rm config} = 1$ within statistical uncertainties. The horizontal line denotes the exact prediction.}
\label{fig:gww_beta_config}
\end{figure*}

Before examining the individual sectors, it is useful to verify that the measured configurational temperature is consistent with the exact Schwinger-Dyson identity. 
Figure~\ref{fig:gww_beta_config} shows the measured configurational temperature in the Gross--Witten--Wadia model for several values of the matrix size and coupling. 
Across the full parameter range investigated, the estimator remains consistent with the exact prediction $\beta_{\rm config} = 1$ within statistical uncertainties.

The agreement is nontrivial because the estimator depends simultaneously on the gradient and Hessian of the action, thereby probing a large number of degrees of freedom in the sampled probability measure. 
The results provide a nontrivial consistency check of both the implementation of the estimator and the Monte Carlo sampling procedure.

Having established the configurational-temperature identity, we now turn to the behavior of the isotropic and anisotropic sectors separately.

\subsection{Gross--Witten--Wadia Model}

We begin with the Gross--Witten--Wadia model, which undergoes a third-order large-$N$ phase transition at the critical coupling $\lambda_c = 1$ \cite{Gross:1980he,Wadia:1980cp}. 

\begin{figure}[htbp]
\centering
\includegraphics[width=14cm]{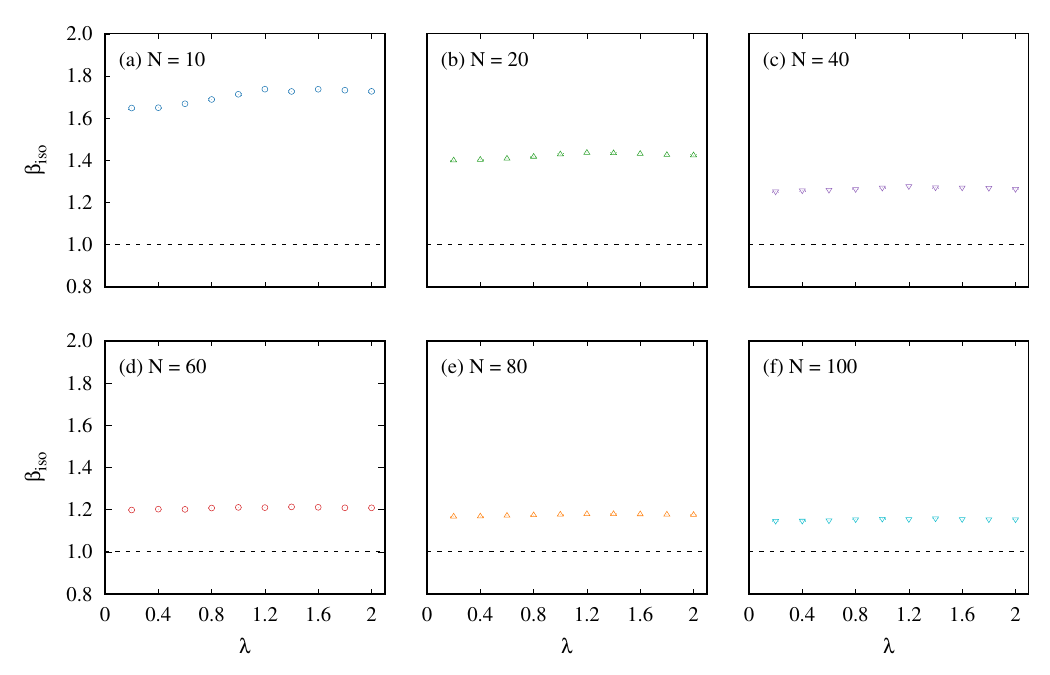}
\caption{The Gross--Witten--Wadia model. Isotropic contribution $\beta_{\rm iso}$ as a function of the coupling $\lambda$ for several matrix sizes. The isotropic sector approaches $\beta_{\rm iso} = 1$ as the matrix size increases, with finite-$N$ corrections becoming progressively smaller. The horizontal line indicates the large-$N$ limit $\beta_{\rm iso} = 1$.}
\label{fig:gww_iso}
\end{figure}

\begin{figure}[htbp]
\centering
\includegraphics[width=14cm]{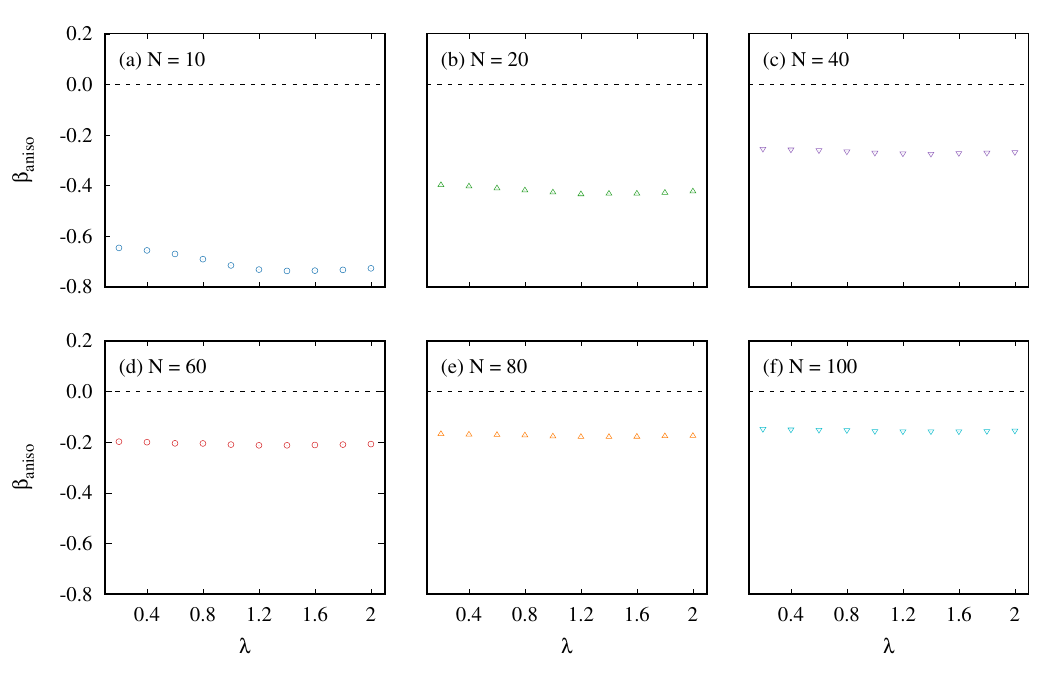}
\caption{The Gross--Witten--Wadia model. Anisotropic contribution $\beta_{\rm aniso}$ as a function of the coupling $\lambda$ for several matrix sizes. The magnitude of the anisotropic sector decreases systematically with increasing $N$, approaching $\beta_{\rm aniso} = 0$ in the large-$N$ limit. The horizontal line indicates $\beta_{\rm aniso} = 0$.}
\label{fig:gww_aniso}
\end{figure}

The isotropic and anisotropic contributions in the Gross--Witten--Wadia model are shown in Figs.~\ref{fig:gww_iso} and \ref{fig:gww_aniso}. 
For all couplings studied, the isotropic sector approaches unity as the matrix size increases, while the anisotropic contribution decreases toward zero,
\begin{equation}
\beta_{\rm iso}\rightarrow 1,
\qquad
\beta_{\rm aniso}\rightarrow 0,
\qquad
N\rightarrow\infty.
\end{equation}

To quantify the approach to the large-$N$ limit, we fit the finite-$N$ corrections using
\begin{equation}
\beta_{\rm iso} - 1 = A_{\rm iso} N^{B_{\rm iso}}, 
\end{equation} 
and 
\begin{equation}
- \beta_{\rm aniso} = A_{\rm aniso} N^{B_{\rm aniso}}. 
\end{equation} 

\begin{figure*}[htbp]
\centering
\includegraphics[width=15cm]{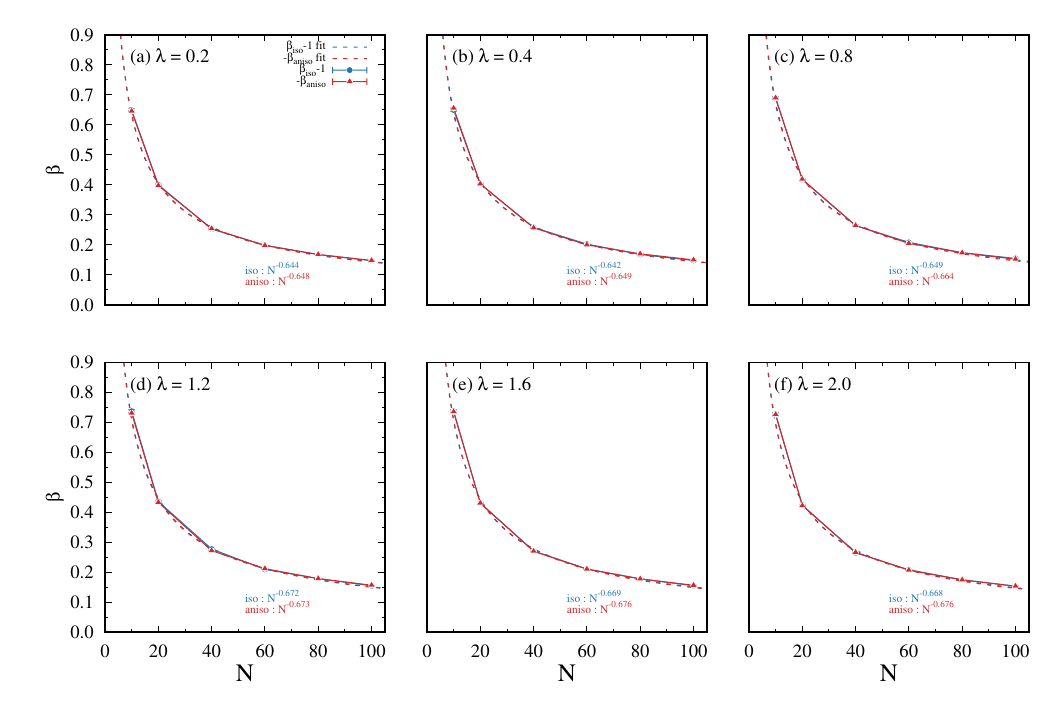}
\caption{The Gross--Witten--Wadia model. Finite-$N$ scaling of the isotropic and anisotropic sectors for couplings $\lambda = 0.2, 0.4, 0.8, 1.2, 1.6$, and $2.0$. The solid lines show fits to the forms $\beta_{\rm iso} - 1 = A_{\rm iso} N^{B_{\rm iso}}$ and $- \beta_{\rm aniso} = A_{\rm aniso} N^{B_{\rm aniso}}$. The isotropic and anisotropic sectors exhibit similar scaling behavior over the range of matrix sizes studied.}
\label{fig:gww_scaling}
\end{figure*}

Representative fits are displayed in Fig.~\ref{fig:gww_scaling}, while the extracted parameters are listed in Table~\ref{tab:gww_scaling}. 

\begin{table}[h]
\begin{center}
\begin{tabular}{c c c c c}
$\lambda$ & $A_{\rm iso}$ & $B_{\rm iso}$ & $A_{\rm aniso}$ & $B_{\rm aniso}$ \\ \hline \hline
$0.2$ & $2.7920(1229)$ & $-0.6442(0119)$ & $2.8184(1152)$ & $-0.6479(0122)$ \\
$0.4$ & $2.7966(0894)$ & $-0.6421(0086)$ & $2.8707(1218)$ & $-0.6494(0127)$ \\
$0.8$ & $2.9788(1624)$ & $-0.6493(0147)$ & $3.1188(1373)$ & $-0.6638(0131)$ \\
$1.2$ & $3.3569(1821)$ & $-0.6722(0146)$ & $3.3364(1984)$ & $-0.6726(0175)$ \\
$1.6$ & $3.2905(2234)$ & $-0.6686(0183)$ & $3.3590(2335)$ & $-0.6759(0204)$ \\
$2.0$ & $3.2270(2417)$ & $-0.6677(0201)$ & $3.3117(2409)$ & $-0.6765(0213)$ \\ \hline
\end{tabular}
\end{center}
\caption{The Gross--Witten--Wadia model. Fit parameters obtained from the finite-$N$ scaling forms $\beta_{\rm iso} - 1 = A_{\rm iso} N^{B_{\rm iso}}$ and $- \beta_{\rm aniso} = A_{\rm aniso} N^{B_{\rm aniso}}$ for several values of the 't Hooft coupling $\lambda$. Numbers in parentheses denote one-standard-deviation uncertainties on the final quoted digits.}
\label{tab:gww_scaling}
\end{table} 

As shown in Table~\ref{tab:gww_scaling}, the exponents extracted from the isotropic and anisotropic sectors are statistically compatible for all couplings considered. 
The fitted exponents cluster around
\begin{equation}
B_{\rm iso}\simeq B_{\rm aniso}\simeq -0.65.
\end{equation}
No strong dependence of the scaling exponent on the coupling is observed over the range of couplings studied.

The amplitudes are also of comparable magnitude. 
Together with the similarity of the scaling exponents, this implies that the leading finite-$N$ corrections satisfy the approximate relation
\begin{equation} 
\beta_{\rm iso} - 1 \simeq - \beta_{\rm aniso}.
\label{eq:GWWmatched} 
\end{equation} 

Over the range of matrix sizes investigated, the leading finite-$N$ corrections are therefore well approximated by
\begin{equation}
\beta_{\rm iso} - 1 \sim - \beta_{\rm aniso} \sim N^{-0.65}.
\end{equation}

\subsection{Double-Well Matrix Model} 

We now turn to the quartic double-well matrix model, which provides a second example of an interacting matrix ensemble with a nontrivial eigenvalue distribution. 

\begin{figure*}[htbp]
\centering
\includegraphics[width=8cm]{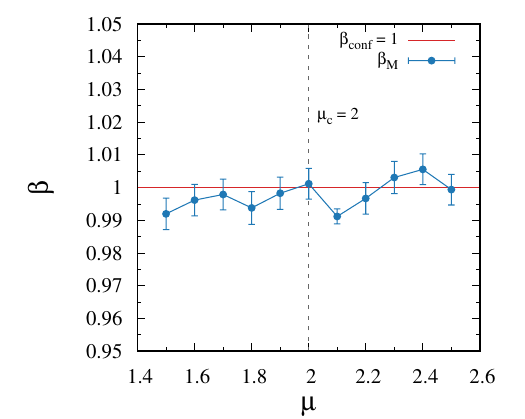}\includegraphics[width=8cm]{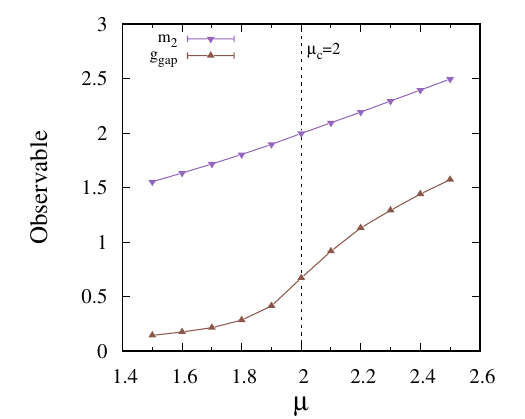}
\caption{
The double-well matrix model. (Left) Measured configurational temperature $\beta_M$ as a function of the coupling $\mu$ for $N = 100$. The estimator remains consistent with the exact Schwinger--Dyson identity $\beta_{\rm config} = 1$ throughout the parameter range studied. The horizontal line denotes the exact prediction. (Right) The observables $\langle m_2\rangle$ and the spectral gap $g_{\rm gap}$ as functions of $\mu$ for $N = 100$, illustrating the behavior of the system across the crossover region.
}
\label{fig:dw_observables}
\end{figure*}

Figure~\ref{fig:dw_observables} shows the measured configurational temperature together with the observables
\begin{equation}
m_2 = \frac1N \left< {\rm Tr} \, M^2 \right>, 
\end{equation}
and the spectral gap $g_{\rm gap}$ as functions of the coupling $\mu$. 
The observables $m_2$ and $g_{\rm gap}$ are included to characterize the evolution of the eigenvalue distribution across the crossover region and to provide context for the configurational-temperature measurements. 
Throughout the interval
\begin{equation}
1.5 \le \mu \le 2.5,
\end{equation}
the configurational temperature remains consistent with the exact prediction $\beta_{\rm config} = 1$.

\begin{figure*}[htbp]
\centering
\includegraphics[width=5cm]{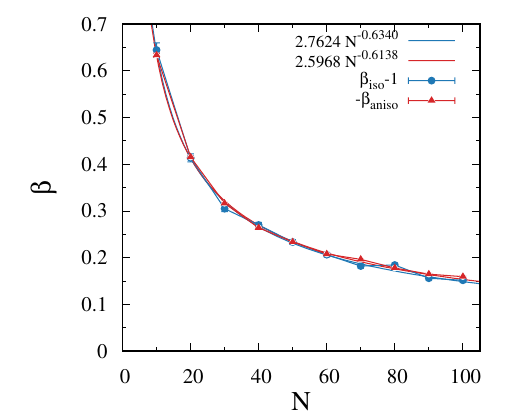}\includegraphics[width=5cm]{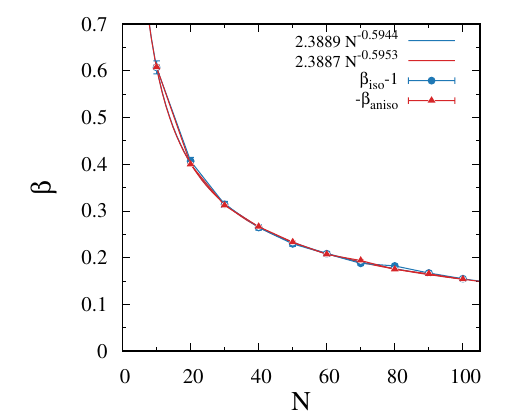}\includegraphics[width=5cm]{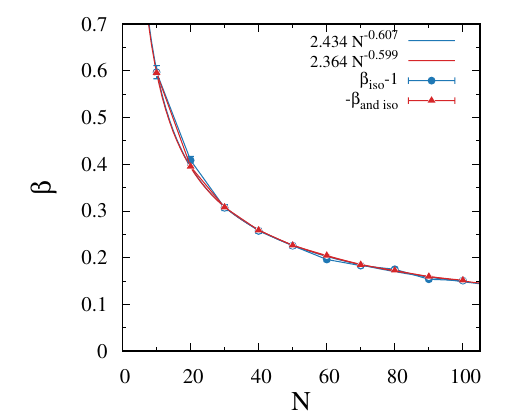}
\caption{
The double-well matrix model. Finite-$N$ scaling of the isotropic and anisotropic sectors for couplings $\mu = 1.5$ (left), $2.0$ (middle), and $2.5$ (right). The solid lines show fits to the forms $\beta_{\rm iso}-1 = A_{\rm iso}N^{B_{\rm iso}}$ and $-\beta_{\rm aniso} = A_{\rm aniso}N^{B_{\rm aniso}}$. The isotropic and anisotropic sectors exhibit similar scaling behavior over the range of matrix sizes studied.
}
\label{fig:dw_scaling}
\end{figure*}

\begin{table}[h]
\begin{center}
\begin{tabular}{ c c c c}
& $\mu = 1.5$ & $\mu = 2.0$ & $\mu = 2.5$ \\ \hline \hline
$A_{\rm iso}$ & $\phantom{-}2.7624(0866)$ & $\phantom{-}2.3889(0353)$ & $\phantom{-}2.4341(0663)$ \\ 
$A_{\rm aniso}$ & $\phantom{-}2.5968(0463)$ & $\phantom{-}2.3887(0226)$ & $\phantom{-}2.3641(0121)$ \\
$B_{\rm iso}$ & $-0.6340(0100)$ & $-0.5944(0046)$ & $-0.6069(0086)$ \\
$B_{\rm aniso}$ & $-0.6138(0056)$ & $-0.5953(0030)$ & $-0.5987(0016)$ \\ \hline  
\end{tabular}
\end{center}
\caption{
The double-well matrix model. Fit parameters obtained from the finite-$N$ scaling forms $\beta_{\rm iso} - 1 = A_{\rm iso} N^{B_{\rm iso}}$ and $- \beta_{\rm aniso} = A_{\rm aniso} N^{B_{\rm aniso}}$ for couplings $\mu = 1.5$, $2.0$, and $2.5$. Numbers in parentheses denote one-standard-deviation uncertainties on the final quoted digits.
}
\label{tab:dw_scaling}
\end{table}

The finite-$N$ behavior of the isotropic and anisotropic sectors is shown in Fig.~\ref{fig:dw_scaling}. As in the Gross--Witten--Wadia model, both sectors are well described by power-law fits of the form $A \, N^B$. The fitted parameters are summarized in Table~\ref{tab:dw_scaling}. 

As shown in Table~\ref{tab:dw_scaling}, the isotropic and anisotropic sectors yield statistically compatible scaling exponents for all values of $\mu$ considered. 
The fitted exponents cluster around
\begin{equation} 
B_{\rm iso} \simeq B_{\rm aniso} \simeq - 0.6. 
\end{equation}
No significant dependence of the scaling exponent on the coupling $\mu$ is observed within the statistical precision of the present study.

The amplitudes are also similar in magnitude, leading again to the approximate relation
\begin{equation} 
\beta_{\rm iso} - 1 \simeq - \beta_{\rm aniso}. 
\label{eq:DWmatched} 
\end{equation} 

The observed finite-$N$ corrections are therefore characterized by 
\begin{equation} 
\beta_{\rm iso} - 1 \sim - \beta_{\rm aniso} \sim N^{- 0.6}. 
\end{equation}

\subsection{Comparison of Interacting Models}

The Gross--Witten--Wadia and double-well matrix models exhibit several common features.

First, in both models, the configurational-temperature estimator reproduces the exact Schwinger--Dyson identity within statistical uncertainties over the full parameter range investigated.

Second, the isotropic and anisotropic sectors exhibit finite-$N$ corrections with compatible scaling exponents and amplitudes of comparable magnitude.

Neither model exhibits a strong dependence of the scaling exponent on the coupling over the parameter ranges investigated.

As a consequence, both systems satisfy the approximate relation
\begin{equation} 
\beta_{\rm iso} - 1 \simeq - \beta_{\rm aniso}, 
\end{equation} 
over the range of matrix sizes considered in this work.

The two interacting matrix models exhibit broadly similar finite-$N$ scaling. 
The fitted exponents cluster around $-0.65$ in the Gross--Witten--Wadia model and around $-0.6$ in the double-well model, with no strong dependence on the coupling observed in either case.

These results provide a useful reference point for the Gaussian random-matrix ensembles studied in the following section.

\section{Configurational Temperature in Gaussian Ensembles}
\label{sec:GaussianResults}

The Gaussian random-matrix ensembles provide useful reference systems for interpreting the behavior observed in the interacting matrix models. 
Their eigenvalue statistics and dynamics have been extensively studied beginning with Dyson's Brownian-motion formulation of random matrices~\cite{Dyson:1962brm}; see also Ref.~\cite{mehta2004random, forrester2010log}.
 
Although the equilibrium measures of the GOE, GUE, and GSE are considerably simpler than those of the Gross--Witten--Wadia and double-well models, they nevertheless contain nontrivial eigenvalue correlations generated by the Vandermonde determinant. 

In this section, we examine the configurational-temperature estimator and its isotropic and anisotropic sectors in the three classical Gaussian ensembles. 
Our primary goal is to compare their finite-$N$ behavior with that observed in the interacting matrix models. 

\subsection{Gaussian Orthogonal Ensemble} 

\begin{figure*}[htbp]
\centering
\includegraphics[width=8cm]{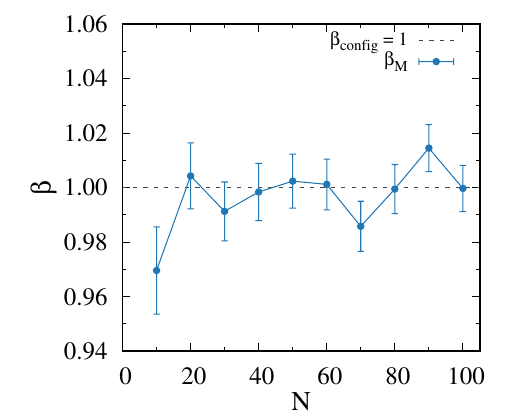}\includegraphics[width=8cm]{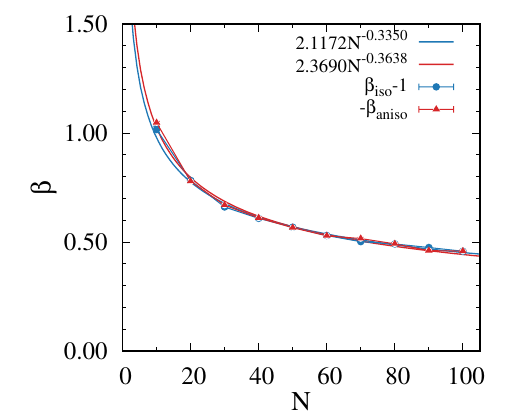}
\caption{
The Gaussian Orthogonal Ensemble. (Left) Measured configurational temperature $\beta_M$ as a function of the matrix size $N$. The estimator remains consistent with the exact Schwinger--Dyson identity $\beta_{\rm config} = 1$ within statistical uncertainties. The horizontal line denotes the exact prediction. (Right) Finite-$N$ behavior of the isotropic and anisotropic sectors. Both $\beta_{\rm iso} - 1$ and $- \beta_{\rm aniso}$ decrease with increasing $N$ and exhibit similar scaling behavior over the range of matrix sizes studied.
}
\label{fig:goe_beta}
\end{figure*}

The numerical results for the Gaussian Orthogonal Ensemble are shown in Fig.~\ref{fig:goe_beta}, while the measured observables and scaling parameters are summarized in Tables~\ref{tab:GOE} and \ref{tab:GOE_scaling}. 

\begin{table}[h]
\begin{center}
\begin{tabular}{ l c r c}
  $N$ & $\beta_M$ & $\beta_{\rm iso} - 1$ & $-\beta_{\rm aniso}$ \\ \hline \hline
  10 & 0.9696(160) & 1.0172(145) & 1.0476(63) \\
  20 & 1.0043(121) & 0.7833(084) & 0.7790(55) \\
  30 & 0.9913(108) & 0.6614(066) & 0.6701(56) \\
  40 & 0.9984(105) & 0.6095(060) & 0.6111(58) \\
  50 & 1.0024(099) & 0.5690(052) & 0.5666(57) \\
  60 & 1.0012(093) & 0.5304(047) & 0.5292(55) \\
  70 & 0.9858(092) & 0.5024(044) & 0.5166(57) \\
  80 & 0.9995(090) & 0.4917(041) & 0.4922(56) \\
  90 & 1.0145(086) & 0.4758(039) & 0.4613(54) \\
  100 & 0.9997(085) & 0.4584(037) & 0.4587(55) \\ \hline
\end{tabular}
\end{center}
\caption{
The Gaussian Orthogonal Ensemble. Measured configurational temperature $\beta_M$ together with the isotropic and anisotropic contributions for matrix sizes $10 \leq N \leq 100$. Numbers in parentheses denote one-standard-deviation statistical uncertainties on the final quoted digits.
}
\label{tab:GOE}
\end{table}

\begin{table}[h]
\begin{center}
\begin{tabular}{ c c c c}
$A_{\rm iso}$ & $A_{\rm aniso}$ & $B_{\rm iso}$ &  $B_{\rm aniso}$ \\ \hline \hline  
$2.1172(0787)$ & $2.3690(783)$ & $-0.3350(92)$ &  $-0.3638(93)$ \\ \hline
\end{tabular}
\end{center}
\caption{
The Gaussian Orthogonal Ensemble. Fit parameters obtained from the finite-$N$ scaling forms $\beta_{\rm iso} - 1 = A_{\rm iso} N^{B_{\rm iso}}$ and $- \beta_{\rm aniso} = A_{\rm aniso} N^{B_{\rm aniso}}$. Numbers in parentheses denote one-standard-deviation uncertainties on the final quoted digits.
}
\label{tab:GOE_scaling}
\end{table}

The configurational temperature remains consistent with the exact identity $\beta_{\rm config} = 1$ throughout the range of matrix sizes investigated. 
This provides a nontrivial check of the estimator in an ensemble whose probability measure is generated entirely by the Gaussian potential and Vandermonde repulsion.

The isotropic and anisotropic sectors exhibit finite-$N$ corrections of comparable magnitude. 
Both sectors decrease as the matrix size increases and are well described by power-law scaling over the range of matrix sizes studied.

Fitting the data to power-law forms yields
\begin{equation}
B_{\rm iso} = - 0.3350(92), \qquad B_{\rm aniso} = - 0.3638(93). 
\end{equation}

The exponents are noticeably smaller in magnitude than those observed in the interacting matrix models, indicating a slower approach to the large-$N$ limit in the GOE ensemble.

\subsection{Gaussian Unitary Ensemble}

\begin{figure*}[htbp]
\centering
\includegraphics[width=8cm]{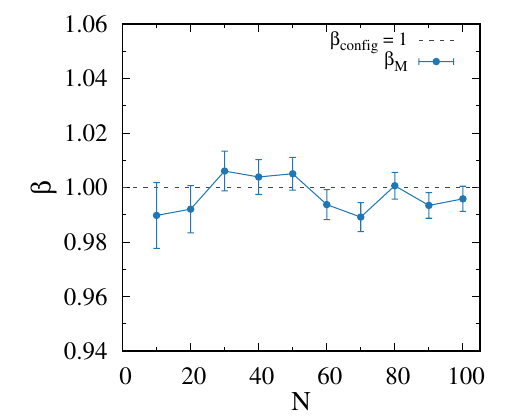}\includegraphics[width=8cm]{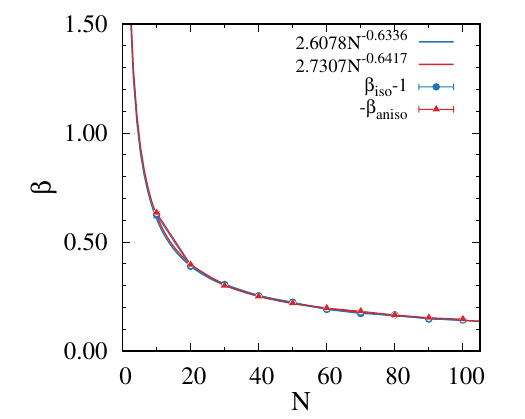}
\caption{The Gaussian Unitary Ensemble. (Left) Measured configurational temperature $\beta_M$ as a function of the matrix size $N$. The estimator remains consistent with the exact Schwinger--Dyson identity $\beta_{\rm config} = 1$ within statistical uncertainties. The horizontal line denotes the exact prediction. (Right) Finite-$N$ behavior of the isotropic and anisotropic sectors. The quantities $\beta_{\rm iso} - 1$ and $- \beta_{\rm aniso}$ decrease with increasing $N$ and exhibit closely similar scaling over the range of matrix sizes studied.
}
\label{fig:gue_beta}
\end{figure*}

Results for the Gaussian Unitary Ensemble are shown in Fig.~\ref{fig:gue_beta}, with numerical values listed in Tables~\ref{tab:GUE} and \ref{tab:GUE_scaling}.
This confirms the configurational-temperature identity in a second Gaussian ensemble with a distinct eigenvalue measure.

\begin{table}[h]
\begin{center}
\begin{tabular}{ l c r c}
  $N$ & $\beta_M$ & $\beta_{\rm iso} - 1$ & $-\beta_{\rm aniso}$ \\ \hline \hline
  10 & 0.9898(121) & 0.6249(132) & 0.6351(34) \\
  20 & 0.9921(087) & 0.3891(022) & 0.3970(22) \\
  30 & 1.0061(073) & 0.3059(063) & 0.2997(19) \\
  40 & 1.0039(064) & 0.2544(053) & 0.2505(19) \\
  50 & 1.0051(060) & 0.2249(048) & 0.2199(18) \\
  60 & 0.9938(055) & 0.1913(043) & 0.1975(18) \\
  70 & 0.9892(053) & 0.1728(041) & 0.1836(18) \\
  80 & 1.0007(049) & 0.1670(038) & 0.1663(17) \\
  90 & 0.9935(047) & 0.1475(035) & 0.1540(16) \\
  100 & 0.9959(046) & 0.1431(034) & 0.1472(16) \\ \hline
\end{tabular}
\end{center}
\caption{
The Gaussian Unitary Ensemble. Measured configurational temperature $\beta_M$ together with the isotropic and anisotropic contributions for matrix sizes $10 \leq N \leq 100$. Numbers in parentheses denote one-standard-deviation statistical uncertainties on the final quoted digits.
}
\label{tab:GUE}
\end{table}

\begin{table}[h]
\begin{center}
\begin{tabular}{ c c c c}
$A_{\rm iso}$ & $A_{\rm aniso}$ & $B_{\rm iso}$ &  $B_{\rm aniso}$ \\ \hline \hline  
$2.6078(721)$ & $2.7308(868)$ & $-0.6336(82)$ &  $-0.6417(90)$ \\ \hline
\end{tabular}
\end{center}
\caption{
The Gaussian Unitary Ensemble. Fit parameters obtained from the finite-$N$ scaling forms $\beta_{\rm iso}-1 = A_{\rm iso}N^{B_{\rm iso}}$ and $-\beta_{\rm aniso} = A_{\rm aniso} N^{B_{\rm aniso}}$. Numbers in parentheses denote one-standard-deviation uncertainties on the final quoted digits.
}
\label{tab:GUE_scaling}
\end{table}

As in the GOE, the configurational-temperature estimator remains consistent with $\beta_{\rm config} = 1$ within statistical uncertainties for all matrix sizes studied.

The isotropic and anisotropic sectors exhibit finite-$N$ corrections of comparable magnitude and are well described by power-law fits. 
As in the interacting matrix models, the leading corrections satisfy the approximate relation $\beta_{\rm iso}-1 \simeq -\beta_{\rm aniso}$. 
The extracted exponents are
\begin{equation} 
B_{\rm iso} = - 0.6336(82), \qquad B_{\rm aniso} = - 0.6417(90). 
\end{equation}

Within statistical uncertainties, the isotropic and anisotropic sectors exhibit the same scaling behavior. 
The fitted exponent is remarkably close to that observed in the Gross--Witten--Wadia model, despite the substantial differences between the underlying probability measures.

The GUE therefore provides an example in which the isotropic and anisotropic sectors not only have comparable amplitudes, but also exhibit statistically indistinguishable scaling exponents.

\subsection{Gaussian Symplectic Ensemble}

\begin{figure*}[htbp]
\centering
\includegraphics[width=8cm]{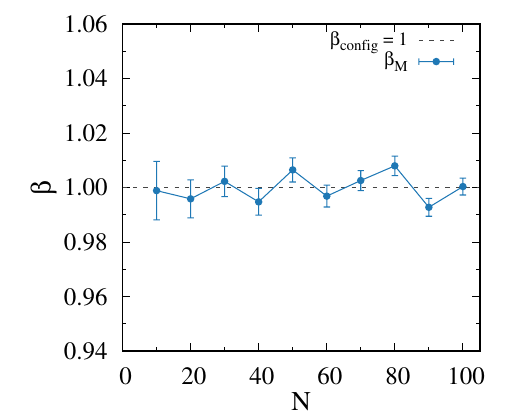}\includegraphics[width=8cm]{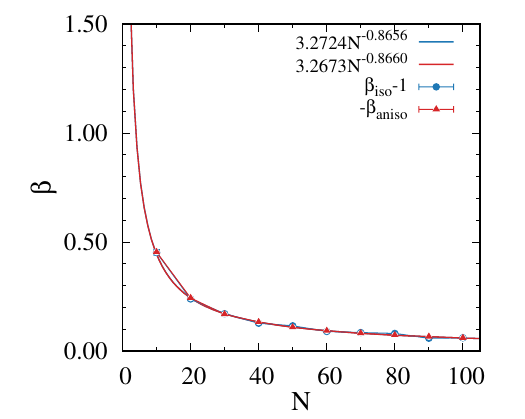}
\caption{
The Gaussian Symplectic Ensemble. (Left) Measured configurational temperature $\beta_M$ as a function of the matrix size $N$. The estimator remains consistent with the exact Schwinger--Dyson identity $\beta_{\rm config} = 1$ within statistical uncertainties. The horizontal line denotes the exact prediction. (Right) Finite-$N$ behavior of the isotropic and anisotropic sectors. The quantities $\beta_{\rm iso} - 1$ and $-\beta_{\rm aniso}$ decrease rapidly with increasing $N$ and exhibit nearly identical scaling over the range of matrix sizes studied.
}
\label{fig:gse_beta}
\end{figure*}

The corresponding results for the Gaussian Symplectic Ensemble are presented in Fig.~\ref{fig:gse_beta}, with measurements and fit parameters given in Tables~\ref{tab:GSE} and \ref{tab:GSE_scaling}.

\begin{table}[h]
\begin{center}
\begin{tabular}{ l c r c}
  $N$ & $\beta_M$ & $\beta_{\rm iso} - 1$ & $-\beta_{\rm aniso}$ \\ \hline \hline
  10 & 0.9989(107) & 0.4533(128) & 0.4544(28) \\
  20 & 0.9959(070) & 0.2397(073) & 0.2438(10) \\
  30 & 1.0023(056) & 0.1720(056) & 0.1697(07) \\
  40 & 0.9948(049) & 0.1284(048) & 0.1335(06) \\
  50 & 1.0065(044) & 0.1163(042) & 0.1099(05) \\
  60 & 0.9969(040) & 0.0905(038) & 0.0936(05) \\
  70 & 1.0026(037) & 0.0855(036) & 0.0828(04) \\
  80 & 1.0080(036) & 0.0819(034) & 0.0738(04) \\
  90 & 0.9928(033) & 0.0597(031) & 0.0669(04) \\
  100 & 1.0004(031) & 0.0618(029) & 0.0613(04) \\ \hline
\end{tabular}
\end{center}
\caption{
The Gaussian Symplectic Ensemble. Measured configurational temperature $\beta_M$ together with the isotropic and anisotropic contributions for matrix sizes $10 \leq N \leq 100$. Numbers in parentheses denote one-standard-deviation statistical uncertainties on the final quoted digits.
}
\label{tab:GSE}
\end{table}

\begin{table}[h]
\begin{center}
\begin{tabular}{ c c c c}
$A_{\rm iso}$ & $A_{\rm aniso}$ & $B_{\rm iso}$ &  $B_{\rm aniso}$ \\ \hline \hline  
$3.2724(2709)$ & $3.2673(0613)$ & $-0.8656(228)$ &  $-0.8660(050)$ \\ \hline
\end{tabular}
\end{center}
\caption{
The Gaussian Symplectic Ensemble. Fit parameters obtained from the finite-$N$ scaling forms $\beta_{\rm iso} - 1 = A_{\rm iso} N^{B_{\rm iso}}$ and $- \beta_{\rm aniso} = A_{\rm aniso}N^{B_{\rm aniso}}$. Numbers in parentheses denote one-standard-deviation uncertainties on the final quoted digits.
}
\label{tab:GSE_scaling}
\end{table}

The configurational-temperature identity remains satisfied within statistical uncertainties, $\beta_{\rm config}=1$, for all matrix sizes studied. 
The isotropic and anisotropic sectors exhibit finite-$N$ corrections of comparable magnitude and are well described by power-law scaling.

The fitted exponents are
\begin{equation}
B_{\rm iso} = - 0.8656(228), \qquad B_{\rm aniso} = - 0.8660(50). 
\end{equation}

Within statistical uncertainties, the isotropic and anisotropic sectors exhibit identical scaling behavior. 
The corresponding exponent is substantially larger in magnitude than those observed in the GOE and GUE, indicating a more rapid approach to the large-$N$ limit.

\subsection{Comparison of Ensembles}

The results obtained for the Gaussian ensembles reveal several common features.

First, all three ensembles satisfy the exact configurational-temperature identity, $\beta_{\rm config} = 1$, within statistical uncertainties.

Second, the isotropic and anisotropic sectors exhibit finite-$N$ corrections of comparable magnitude in every ensemble studied. 
In most cases, the exponents extracted from the two sectors are statistically compatible, while the corresponding amplitudes are of similar size. 
As a result, $\beta_{\rm iso} - 1 \simeq - \beta_{\rm aniso}$ provides a good description of the leading finite-$N$ corrections.

The principal difference between the ensembles lies in the value of the scaling exponent. Table~\ref{tab:all_scalings} summarizes the results obtained in the Gaussian ensembles and interacting matrix models.

\begin{table}[t]
\begin{center}
\begin{tabular}{lcc}
Model & $B_{\rm iso}$ & $B_{\rm aniso}$ \\
\hline\hline
GOE & $-0.3350(92)$ & $-0.3638(93)$ \\
GUE & $-0.6336(82)$ & $-0.6417(90)$ \\
GSE & $-0.8656(228)$ & $-0.8660(50)$ \\
Double Well & $-0.59$ to $-0.63$ & $-0.60$ to $-0.61$ \\
GWW & $-0.64$ to $-0.67$ & $-0.65$ to $-0.68$ \\
\hline
\end{tabular}
\end{center}
\caption{
Summary of finite-$N$ scaling exponents obtained from the fits $\beta_{\rm iso} - 1 = A_{\rm iso} N^{B_{\rm iso}}$ and $- \beta_{\rm aniso} = A_{\rm aniso} N^{B_{\rm aniso}}$. Numbers in parentheses denote one-standard-deviation uncertainties on the final quoted digits.
}
\label{tab:all_scalings}
\end{table}

The most significant variation between ensembles is found in the scaling exponent. 
The GOE exhibits relatively slow finite-$N$ corrections, with exponents near $-0.35$, whereas the GUE and GSE display progressively larger exponents of approximately $-0.64$ and $-0.87$, respectively. 
The interacting matrix models studied here lie between these cases, with exponents near $-0.6$ in the double-well model and near $-0.65$ in the Gross--Witten--Wadia model.

Taken together, these results indicate that the qualitative finite-$N$ structure observed in the interacting matrix models is not unique to those systems. 
In particular, the approximate relation $\beta_{\rm iso}-1 \simeq -\beta_{\rm aniso}$ appears across a broad class of matrix ensembles. 
At the same time, the value of the scaling exponent depends significantly on the underlying probability measure, demonstrating that the detailed approach to the large-$N$ limit is model dependent. 
The numerical results, therefore, suggest that the approximate cancellation between the isotropic and anisotropic sectors is considerably more robust than the associated scaling exponent.

\section{Configurational Temperature as a Monte Carlo Diagnostic}
\label{sec:Diagnostics}

In the previous sections, we studied the configurational-temperature estimator as an observable characterizing the probability measure of matrix models and random-matrix ensembles. In addition to this role, the estimator provides a useful diagnostic for assessing the quality of Monte Carlo simulations.

The usefulness of the configurational temperature follows directly from the exact identity $\beta_{\rm config} = 1$, which must hold for any correctly sampled ensemble. Deviations from unity, therefore, indicate that the sampled distribution differs from the target probability measure.

Unlike conventional observables, which probe particular moments or correlation functions, configurational temperature is derived directly from the probability measure through an exact Schwinger--Dyson identity. 
It therefore provides a complementary test of equilibration and sampling correctness.

In this section, we illustrate these features using simulations of the Gross--Witten--Wadia model.

\subsection{Thermalization of the Probability Measure}

Determining whether a Monte Carlo simulation has reached equilibrium is often a nontrivial task. 
In practice, thermalization is commonly assessed by monitoring one or more physical observables and verifying that they fluctuate around stationary values.

To illustrate the behavior of the configurational-temperature estimator during thermalization, we initialize the Gross--Witten--Wadia model from an ordered configuration,
\begin{equation}
\theta_i = 0, \qquad i = 1, \ldots, N, 
\end{equation}
and monitor both the configurational temperature and the Polyakov loop as functions of Monte Carlo time.

The Polyakov loop is defined by
\begin{equation}
P = \frac1N \sum_{i = 1}^{N} e^{i \theta_i}. 
\end{equation}

\begin{figure}[htbp]
\centering
\includegraphics[width=10cm]{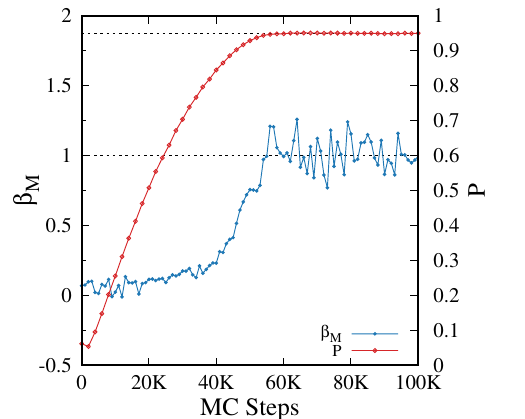}
\caption{
Thermalization in the Gross--Witten--Wadia model. Evolution of the configurational temperature $\beta_M$ and the Polyakov loop as functions of Monte Carlo time. Both observables approach their equilibrium values over comparable Monte Carlo time scales. The horizontal reference lines denote the exact Schwinger--Dyson value $\beta_{\rm config} = 1$ and the measured equilibrium expectation value of the Polyakov loop, $P \simeq 0.95$.
}
\label{fig:thermalization}
\end{figure}

The resulting evolution is shown in Fig.~\ref{fig:thermalization}.

At early Monte Carlo times, both the Polyakov loop and the configurational temperature evolve toward their equilibrium values. 
The Polyakov loop reaches a nearly stationary value relatively quickly, while the configurational temperature approaches the exact Schwinger--Dyson value, $\beta_{\rm config} = 1$, more gradually. 
As the simulation proceeds, the configurational temperature becomes consistent with the exact identity, indicating convergence of the sampled ensemble to the target probability measure. 

These results demonstrate that the configurational temperature provides a direct measure-level test of equilibration based on an exact identity of the probability measure. 
In contrast, conventional observables such as the Polyakov loop must be compared with their equilibrium expectation values, which generally depend on the model and simulation parameters. 
The configurational temperature therefore provides a useful complementary diagnostic for monitoring the approach to equilibrium.

\subsection{Detection of Sampling Distortions}

The configurational temperature is also sensitive to violations of the target probability measure.

To demonstrate this feature, we introduce a bias into the Metropolis proposal according to
\begin{equation}
\theta_i \rightarrow \theta_i + \delta + \mu, 
\end{equation}
where
\begin{equation}
\delta \in [- \epsilon, \epsilon] 
\end{equation}
is a symmetric random increment and $\mu$ is a fixed bias parameter. 
Because the proposal distribution is no longer symmetric, using the standard Metropolis acceptance probability generally violates detailed balance and changes the stationary distribution.

\begin{figure}[htbp]
\centering
\includegraphics[width=8cm]{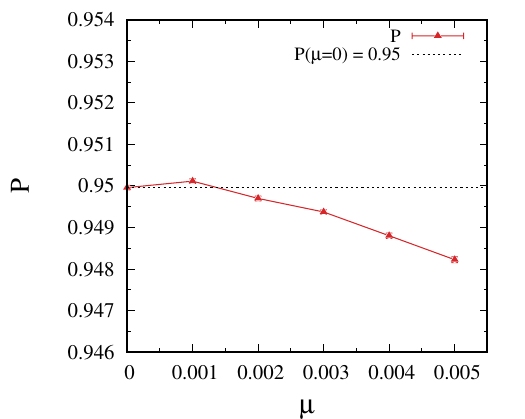}\includegraphics[width=8cm]{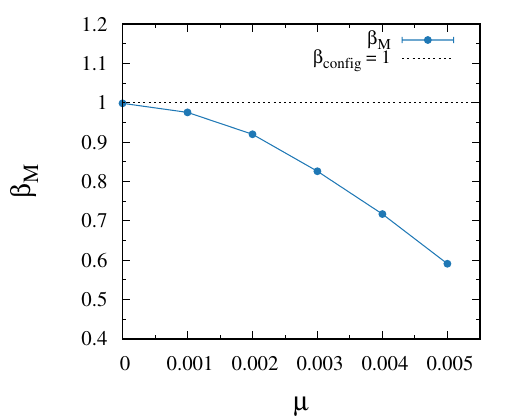}
\caption{
Response of the Polyakov loop and configurational temperature to biased Metropolis updates in the Gross--Witten--Wadia model. (Left) Polyakov loop as a function of the bias parameter $\mu$. (Right) Configurational temperature $\beta_M$ as a function of $\mu$. As the update bias increases, the configurational temperature develops systematic deviations from the exact Schwinger--Dyson prediction $\beta_{\rm config} = 1$, indicating distortions of the sampled probability measure. The horizontal line denotes the exact expectation value $\beta_{\rm config} = 1$.
}
\label{fig:biased_updates}
\end{figure}

Figure~\ref{fig:biased_updates} compares the response of the Polyakov loop and configurational temperature as the bias parameter is varied.

The Polyakov loop exhibits only a modest response to small distortions of the probability measure. By contrast, the configurational temperature exhibits clear, systematic deviations from the exact expectation value $\beta_{\rm config} = 1$. 

As the bias is reduced and correct sampling is restored, the configurational temperature approaches unity, consistent with the recovery of the target probability measure.

The response of the estimator demonstrates its sensitivity to violations of detailed balance and to distortions of the sampled ensemble that may not be immediately apparent in conventional observables.

The examples presented above demonstrate two complementary applications of the configurational-temperature estimator. 
First, the estimator can be used to monitor equilibration by testing whether the sampled ensemble satisfies the exact identity $\beta_{\rm config} = 1$. 
Second, it can detect systematic distortions of the probability measure arising from incorrect update procedures. 
Because these tests probe the probability measure directly rather than individual observables, configurational temperature provides a useful complement to conventional diagnostics in Monte Carlo simulations.

Although the examples presented here are based on the Gross--Witten--Wadia model, the construction is completely general and applies to any system for which the gradient and Hessian of the action can be evaluated. 

\section{Discussion and Outlook}
\label{sec:Discussion}

In this work, we have investigated configurational temperature in a variety of matrix models and random-matrix ensembles. 
The estimator is defined through an exact Schwinger--Dyson identity and may be expressed entirely in terms of the gradient and Hessian of the action. 
This representation admits a simple decomposition into isotropic and anisotropic sectors,
\begin{equation}
\beta_{\rm config} = \beta_{\rm iso} + \beta_{\rm aniso},
\end{equation}
whose sum satisfies the exact identity
\begin{equation}
\beta_{\rm config} = 1
\end{equation}
for any correctly sampled ensemble.

Our numerical results reveal a common qualitative structure across all systems studied. In the Gross--Witten--Wadia model, the quartic double-well matrix model, and the Gaussian Orthogonal, Unitary, and Symplectic Ensembles, the isotropic and anisotropic sectors exhibit finite-$N$ corrections of comparable magnitude. 
In each case, the leading corrections approximately satisfy
\begin{equation}
\beta_{\rm iso} - 1 \simeq - \beta_{\rm aniso}.
\end{equation}
That is, cancellations happen between the two sectors while preserving the exact configurational temperature identity.

Although this cancellation pattern appears to be robust, the associated scaling exponents are not universal. 
The Gross--Witten--Wadia and double-well matrix models yield exponents close to $-0.65$ and $-0.6$, respectively, while the Gaussian ensembles display a broader range extending from approximately $-0.35$ in the GOE to approximately $-0.87$ in the GSE. 
These results indicate that the qualitative structure of the estimator is mostly independent of the detailed form of the probability measure. 
In contrast, the quantitative approach to the large-$N$ limit retains information about the underlying ensemble.

The configurational-temperature estimator also proves useful as a diagnostic of Monte Carlo simulations. 
Because it follows from an exact identity of the target probability measure, it provides a direct measure-level test of sampling correctness. 
The thermalization study shows that conventional observables may appear to be equilibrated even when the probability measure has not yet fully converged. 
Similarly, biased update procedures that violate detailed balance generate systematic deviations of the configurational temperature from its exact expectation value. 
These examples illustrate how the estimator can complement more traditional diagnostics based on physical observables. 

The present work suggests several directions for future investigation. 
It would be interesting to analytically understand the origin of the observed finite-$N$ scaling exponents and the approximate cancellation between the isotropic and anisotropic sectors. 
Another extension would be to study configurational temperature in matrix quantum mechanics, lattice field theories, and systems with complex actions, where reliable sampling-correctness diagnostics are difficult to construct. 
It would also be useful to explore whether related Schwinger--Dyson observables can provide additional geometric information about the probability measure beyond that contained in the configurational temperature itself. 

More broadly, the results presented here demonstrate that configurational temperature is not only an exact observable defined by an exact Schwinger--Dyson identity, but also a practical tool for probing finite-$N$ structure and diagnosing numerical simulations. 
We expect that these features will make it useful in a wider range of applications involving Monte Carlo sampling and matrix-model dynamics. 

\acknowledgments 

We extend our gratitude to Navdeep Singh Dhindsa, Vamika Longia, Michael Mandl, and Piyush Kumar for their invaluable discussions. 
The work of A.J. was supported in part by a Start-up Research Grant from the University of the Witwatersrand. 

\bibliographystyle{JHEP}
\bibliography{biblio.bib}
\end{document}